\documentclass[preprint,12pt]{elsarticle}%
\usepackage{amssymb}
\usepackage{amsfonts}
\usepackage{amsmath}
\usepackage{graphicx}%
\providecommand{\U}[1]{\protect\rule{.1in}{.1in}}
\graphicspath{{E:/dottorato/Coulomb_damper/Testo/Paper/Immagini/}}
\journal{journal}

\begin{document}
%
\begin{frontmatter}%


%

\title{On the Afferrante-Carbone theory of ultratough peeling}%

%

\author{M. Ciavarella}%
%

\address
{Politecnico di BARI. DMMM department. Viale Gentile 182, 70126 Bari. Mciava@poliba.it}%
%

\begin{abstract}%

In an interesting theory of ultratough peeling of an elastic tape from a
viscoelastic substrate, Afferrante and Carbone (2016) find that, in contrast
to the classic elastic Kendall's theory, there are conditions for which the
load for steady state peeling could be arbitrarily large in steady state
peeling, at low angles of peeling - what they call "ultratough" peeling. It is
here shown in fact that this occurs near critical speeds where the elastic
energy term of Kendall's equation is balanced by the viscoelastic dissipation.
Surprisingly, this seems to lead to toughness enhancement higher than the
limit value observed in a very large crack in a infinite viscoelastic body,
possibly even considering a limit on the stress transmitted. Kendall's
experiments in turn had considered viscoelastic tapes (rather than
substrates), and his viscoelastic findinds seem to lead to a much simpler
picture. The Afferrante-Carbone theory suggests the viscoelastic effect to be
an \textit{on-off mechanism}, since for large angles of peeling it is almost
insignificant, while only below a certain threshold, this "ultratough" peeling
seems to appear. Experimental verification would be most useful.%

\end{abstract}%

%

\end{frontmatter}%



\section{Introduction}

Afferrante and Carbone ([1], AC in the following) have given an interesting
and very simple extension of the celebrated Kendall solution for the peeling
of a tape from an elastic (or rigid) substrate (Kendall, [2]). The importance
of the application is evident from pressure-sensitive adhesives to biological
attachments in many insects, spiders and lizards or geckos, and we can refer
to the reference list in AC or in the review of [3-4]. There is relatively
extensive literature on crack propagation in viscoelastic materials theories
and extensive measurements ([5-9]) suggesting steady state subcritical crack
propagation occurs with an enhanced work of adhesion $G$ obtained as the
product of adiabatic value $G_{0}$ and a function of velocity of peeling of
the crack line and temperature, namely the Gent-Schultz law $\ $%
\begin{equation}
\frac{G}{G_{0}}=1+\left(  \frac{v}{v_{0}}\right)  ^{n} \label{wvisco}%
\end{equation}
where $v_{0}=\left(  ka_{T}^{n}\right)  ^{-1}$ and $k,n$ are (supposed to be)
constants of the material, with $0<n<1$ and $a_{T}$ is the WLF factor to
translate results at various temperatures $T$ (Williams, Landel \& Ferry,
[11]). There was originally a difficulty in dealing with viscoelastic cracks
([12-13]) which was solved by [9,10,14], with Barenblatt or Maugis-Dugdale
cohesive models zones, which predict the speed dependence of the Gent-Schultz
law, but with a limit being given by at very large speeds (elastic fracture)
\begin{equation}
\frac{G_{\infty}}{G_{0}}=\frac{E_{\infty}}{E_{0}} \label{maxenhancement}%
\end{equation}
for a material having a relaxed modulus $E_{0}$ and an instantaneous modulus
$E_{\infty}$. de Gennes [15] in particular suggested a "viscoelastic trumpet"
crack model which, when writing the LEFM (Linear Elastic Fracture Mechanics)
condition on both the inner region for the stress intensity factor $K$ as
$K^{2}=G_{0}E_{\infty}$ and in the outside region $K^{2}=G_{\infty}E_{0}$,
leads to a maximum apparent adhesion/toughness enhancement of
(\ref{maxenhancement}). Notice that this can be possibly of 3 or 4 orders of
magnitude ([6]).

The AC theory deals with peeling a tape as in Fig.1, where an angle $\theta$
is referred with respect to the substrate line, and leads to a quite simple
exact solution. The perhaps strong assumption made to obtain such a simple
solution was that the stress (both pressure and shear) is constant on a strip
of dimension $L$ equal to the thickness of the tape, where they also assume
the stress is constant (and simply given by equilibrium with the applied
load), so they can compute the dissipation in the steady state propagation at
velocity $v$. We shall discuss some interesting aspects in the "ultratough
peeling" AC theory.

\begin{center}%
\begin{tabular}
[c]{l}%
\centering\includegraphics[height=65mm]{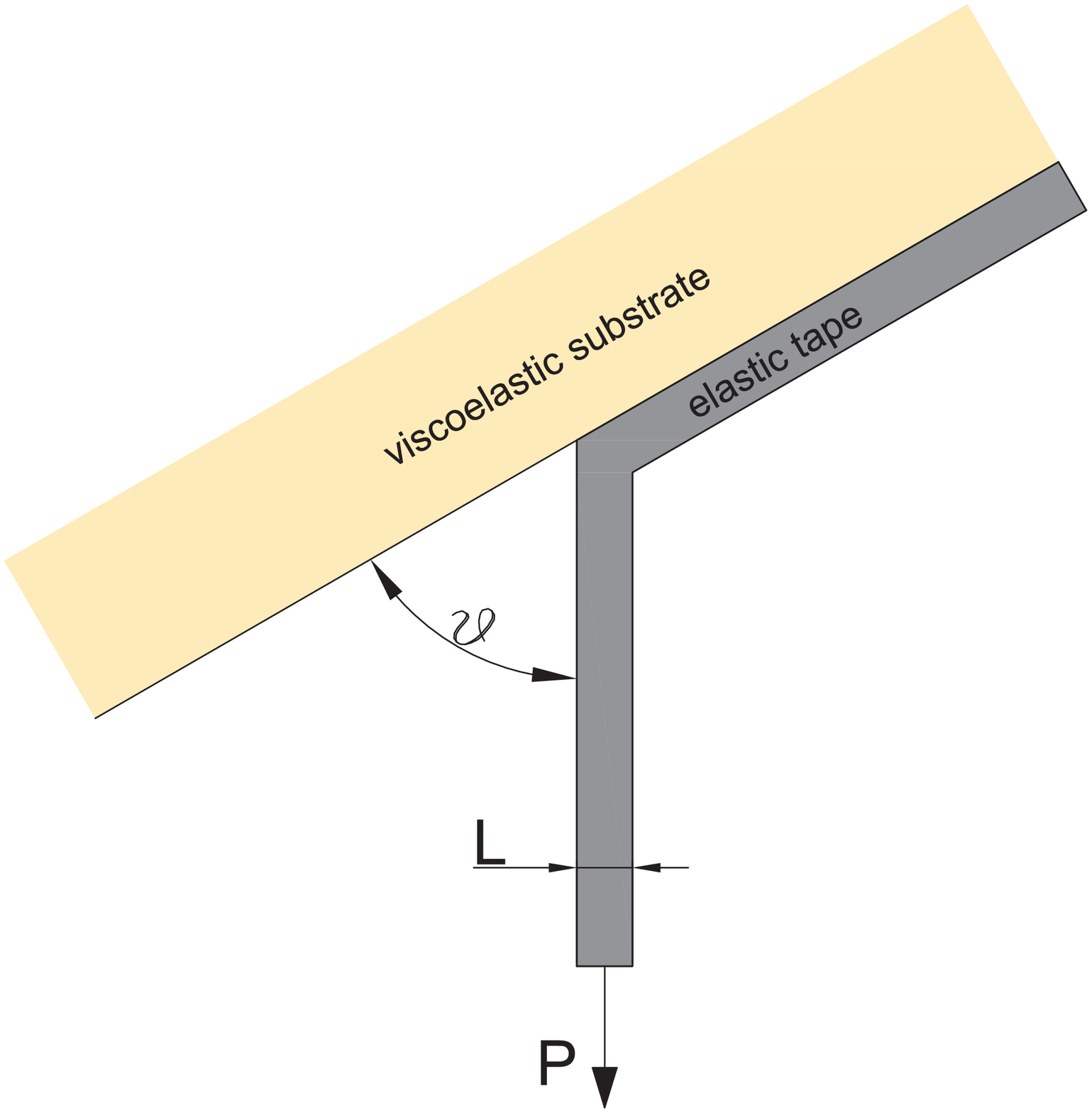}
\end{tabular}

Fig.1 - Geometry of the peeling of an elastic tape from a viscoelastic
substrate of the AC theory.
\end{center}

\section{Afferrante-Carbone theory}

Considering the general case of an elastic tape of modulus $E$, and a
viscoelastic substrate with a single relaxation time $\tau_{0}$, their eqt.14
gives
\begin{equation}
\frac{G_{0}}{EL}=\left[  \frac{1}{2}-\frac{E}{E_{0}}f_{v}\left(  \frac
{v\tau_{0}}{L}\right)  \right]  \left(  \frac{P}{EL}\right)  ^{2}+\left(
\frac{P}{EL}\right)  \left(  1-\cos\theta\right)  \label{1}%
\end{equation}
where $P$ is load per unit out-of-plane thickness, and $L$ is the thickness of
the tape, and finally $f_{v}\left(  \frac{v\tau_{0}}{L}\right)  $ is a
function (see eqt.15 in AC) of exponential integrals which is zero at both
zero speed and infinite speed, and otherwise is positive and has a maximum
which depends weakly on $\frac{E_{\infty}}{E_{0}}$ if $\frac{E_{\infty}}%
{E_{0}}>15$. .

Eqt.(\ref{1}) is a quadratic equation for the load which has obviously real
solutions only in certain ranges for
\begin{equation}
\Delta=\left(  1-\cos\theta\right)  ^{2}+4\frac{G_{0}}{EL}\left[  \frac{1}%
{2}-\frac{E}{E_{0}}f_{v}\left(  \frac{v\tau_{0}}{L}\right)  \right]  >0
\label{delta}%
\end{equation}
and naturally we have to take only positive real roots
\begin{equation}
\left(  \frac{P}{EL}\right)  _{1,2}=\frac{-\left(  1-\cos\theta\right)
\pm\sqrt{\Delta}}{2\left[  \frac{1}{2}-\frac{E}{E_{0}}f_{v}\left(  \frac
{v\tau_{0}}{L}\right)  \right]  }>0\text{ when }\Delta>0 \label{solutions}%
\end{equation}

For $v=0$ or $v=\infty$, the solution returns to well known solution of
peeling of Rivlin [16] for rigid tape or Kendall [2] for elastic tape,
depending on the tape elastic modulus, but this is obvious since there is no
viscoelastic dissipation in these limit cases.

The effect of viscoelasticity at intermediate speeds is contained in the
effective toughness written from (\ref{1}) as
\begin{equation}
\frac{G}{G_{0}}=1+\frac{1}{G_{0}/\left(  EL\right)  }\frac{E}{E_{0}}%
f_{v}\left(  \frac{v\tau_{0}}{L}\right)  \left(  \frac{P}{EL}\right)  ^{2}
\label{2}%
\end{equation}
where $P$ is the solution of (\ref{1}) at the equilibrium condition.

AC find that some ranges of the peeling angle and material combinations exist
for which \textit{"steady state peeling may occur for arbitrarily large values
of the applied force"}, which is what they call "ultratough peeling". In fact,
we find that this is not entirely true, as there is a limit load and this is
not infinite, although it is much larger than for other angles of peeling.
Above this limit load, we expect the system to turn unstable, and the
delamination grow at much higher speeds.

We consider the case $\frac{E_{\infty}}{E_{0}}=10$, $\frac{E}{E_{0}}=1.5$ and
$\frac{G_{0}}{EL}=10^{-4}$ which is also a case considered in the AC\ paper
for their illustrative purposes. We report in Fig.2a the lower of the two
solutions of the load. We find that this corresponds to a toughness
enhancement larger than the infinite system one (in the figure is appears as
$G_{\max}/G_{0}\simeq40$ but a closer look revails that in fact $G/G_{0}%
\simeq140$). This is certainly bigger than $\frac{E_{\infty}}{E_{0}}=10$.
Although there is no rigorous proof that the infinite crack in the infinite
system has the largest possible enhancement, this sounds physically quite
intuitive since a finite system has simply a smaller region which can
dissipate, and is also suggested by theories about crack in finite size
systems like de Gennes [15]. Naturally, there is no attempt to model a true
crack singularity in the AC theory, and this should be further investigated.

There seems to be a dramatic effect of the peeling angle in Fig.2, since for
$\theta>\theta_{th}$ the toughness enhancement is in fact \textit{almost
insignificant} at all speeds, and nothing remotely close to the order we
expect for a large crack in infinite size specimen, that is $E_{\infty}/E_{0}%
$, and see the orders of magnitude difference in Fig.2a, so one could use the
Kendall solution. The viscoelastic effect really seems to be an \textit{on-off
effect} depending on the angle.

Also, as AC say "\textit{Notice equilibrium at higher loads could be, in
principle, established provided the system is forced to move on the upper
branch of the curve.}" And these high low solutions are reported in Fig.2b
where it is found that they correspond to a toughness enhancement of orders of
magnitude larger than the infinite system one. These solutions look really unrealistic.

\bigskip

\begin{center}%
\begin{tabular}
[c]{l}%
\centering\includegraphics[height=65mm]{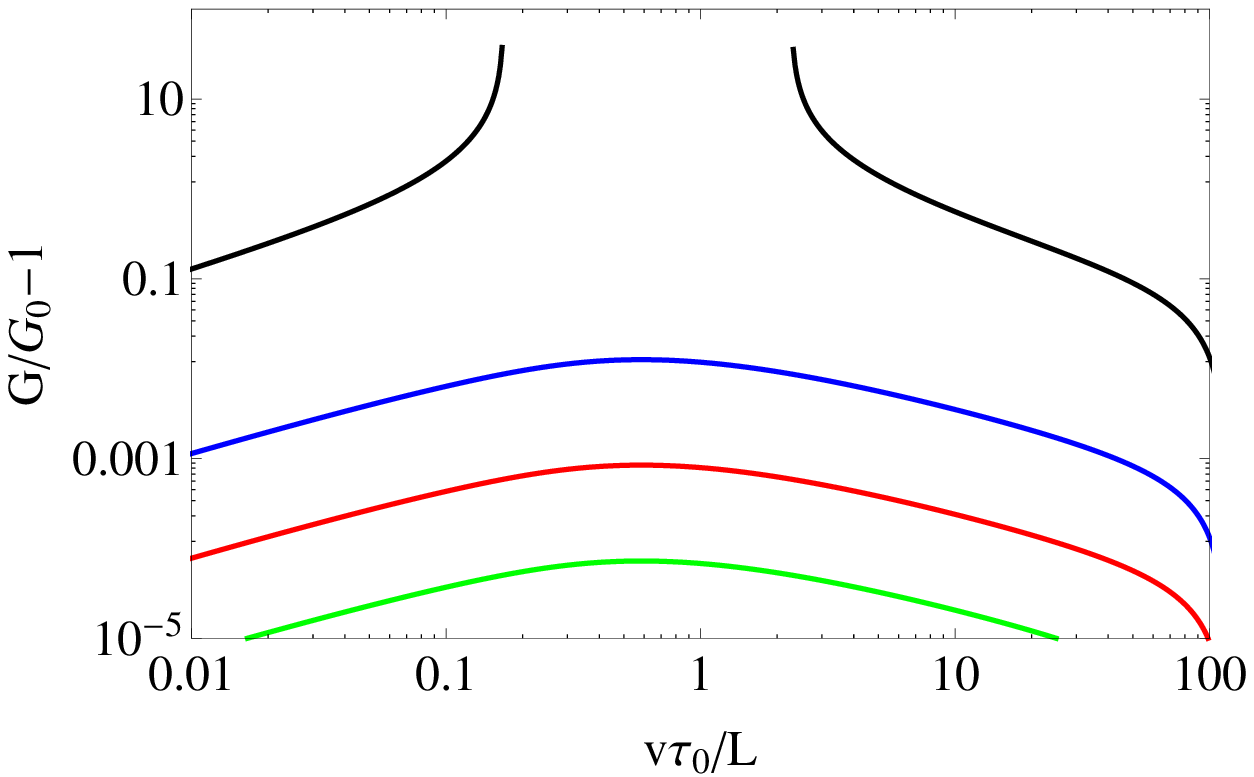}(a)
\end{tabular}

\begin{tabular}
[c]{l}%
\centering\includegraphics[height=65mm]{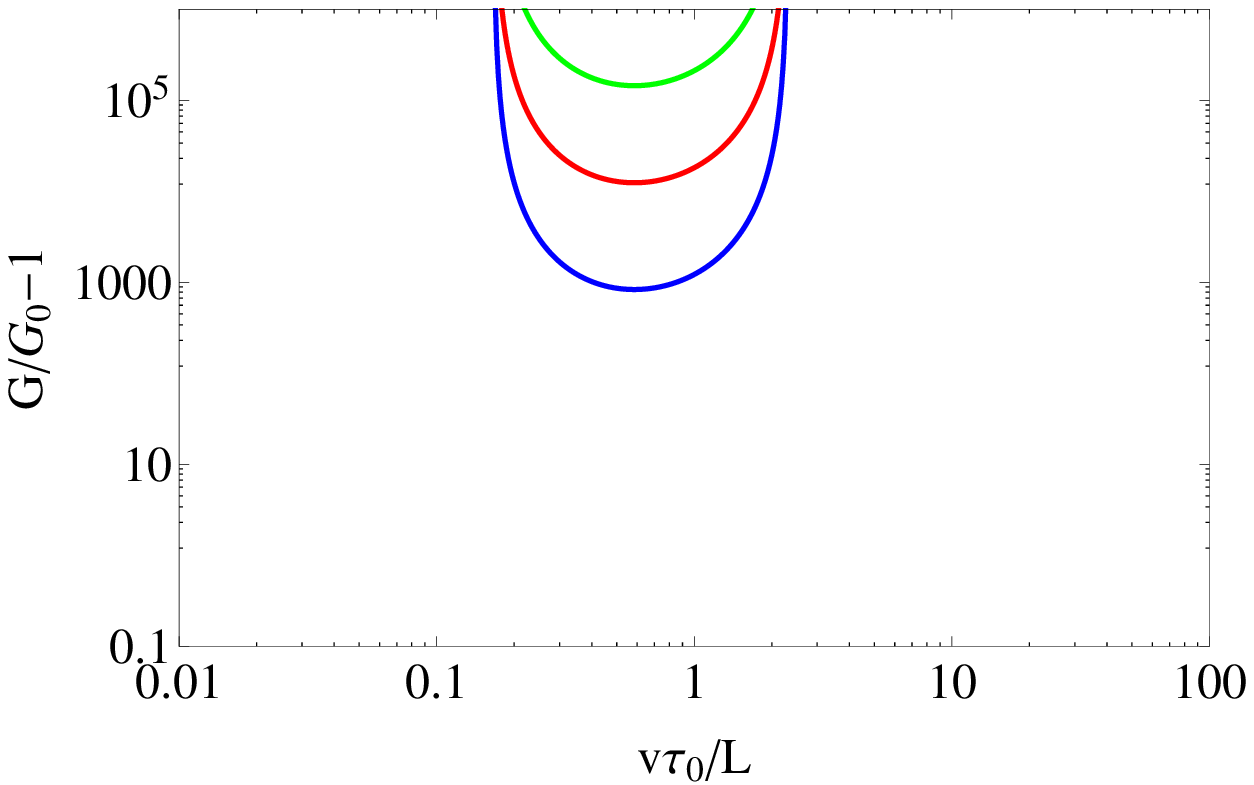}(b)
\end{tabular}

\end{center}

Fig.2 - Enhancement $\left(  G/G_{0}-1\right)  $ for the AC peeling theory
taking as function of the peeling velocity $v$ (a) low load solution (b) large
load solution. Case $\frac{E}{E_{0}}=1.5$. Peeling angles $\theta=\pi
/64,\pi/8,\pi/4,\pi/2$ (black, blue, red, green colors). $\frac{G_{0}}%
{EL}=10^{-4}$, $\frac{E_{\infty}}{E_{0}}=20$.

\bigskip AC find the "ultratough" enhancement when $E/E_{0}$ is not too large.
However, the exact threshold depends on the peeling angle. For $E/E_{0}=1.5$,
ultratough peeling will occur for $\theta<\theta_{th}=0.14rad$, but for higher
$E/E_{0}>1.5$, $\theta_{th}$ will increase and the range will be larger.

\section{\bigskip Limits of the solutions on physical grounds}

\bigskip We could disregard the very large load solutions by adding a cutoff
on the force on physical grounds, as a limit to the stress which can be
carried by the tape\ (or the substrate): as discussed by Kendall (1975) for
elastomers or polymer materials can be of the order of the elastic modulus,
$\sigma_{c}=E$, where in turn $E$ is of the order of $E_{0}$ and $E_{\infty
}>>E_{0}$ means this condition is also an even stronger limit for the
substrate: this results in
\begin{equation}
\frac{P}{EL}<1
\end{equation}

This however will not limit much the maximum amplification. Indeed, taking
$\frac{E}{E_{0}}f_{v}\left(  \frac{v\tau_{0}}{L}\right)  \simeq1$, $\left(
\frac{P}{EL}\right)  ^{2}\simeq1$, i.e. considering the limit on the cohesive
stress (for the higher of the two solution in the quadratic equation)
\begin{equation}
\left(  \frac{G}{G_{0}}\right)  _{\max,1}\simeq\frac{1}{G_{0}/\left(
EL\right)  }%
\end{equation}
(where the subscript "1" stands for the higher of the two solutions of the
quadratic equation), which can of course go to infinity even after the regularization.

\section{Discussion}

Even looking at the smaller of the two solutions of the quadratic equation,
this large amplifications occur when: (i) the coefficient of the quadratic
equation of the AC\ theory becomes nearly zero i.e.
\begin{equation}
\frac{E}{E_{0}}f_{v}\left(  \frac{v\tau_{0}}{L}\right)  \simeq\frac{1}{2}
\label{condition}%
\end{equation}
and (ii) the angle of peeling is low, which makes also the coefficient of the
linear term of the quadratic equation close to zero. Under these conditions,
physically (i) suggests we are considering the case where viscoelastic
dissipation nearly exactly cancels the elastic energy terms of the Kendall
equation, so that one finds the Rivlin equation load
\begin{equation}
P=\frac{G_{0}}{\left(  1-\cos\theta\right)  }%
\end{equation}
and this leads to the toughness enhancement
\begin{equation}
\frac{G}{G_{0}}=1+\frac{E}{E_{0}}f_{v}\left(  \frac{v\tau_{0}}{L}\right)
\frac{G_{0}}{EL}\frac{1}{\left(  1-\cos\theta\right)  ^{2}}%
\end{equation}
which at $\theta=\frac{\pi}{64}rad$, $\frac{E}{E_{0}}f_{v}\left(  \frac
{v\tau_{0}}{L}\right)  \simeq1/2$ and $\frac{G_{0}}{EL}=10^{-4}$ is of the
order of the $\frac{G}{G_{0}}\simeq140\ $we are observing in Fig.2a. Hence,
(ii) suggests that the Rivlin load grows unbounded. Again, assuming the limit
$\frac{P}{EL}=1$ , the Rivlin equation corresponds to the limit $\frac{G_{0}%
}{EL}=\left(  1-\cos\theta\right)  $ and hence the maximum toughness
enhancement at the critical velocities where $\frac{E}{E_{0}}f_{v}\left(
\frac{v\tau_{0}}{L}\right)  \simeq1/2$ is
\begin{equation}
\left(  \frac{G}{G_{0}}\right)  _{\max,2}=1+\frac{1}{2}\frac{1}{1-\cos\theta}%
\end{equation}
(where the subscript "2" stands for the lower of the two solutions of the
quadratic equation) and hence even this can grow still \textit{unbounded} even
if we limit the transmitted stress.

\subsection{Kendall's theory and experiments}

Let us return to the classical elastic Kendall's solution and experiments [2].
At large peeling angles, Kendall did not need to modify the Rivlin solution
("no extension" theory), and since the tape in Kendall's experiment is
viscoelastic, Kendall measured from the Rivlin solution the toughness as \ a
function of speed
\begin{equation}
P_{Rivlin}\left(  v\right)  =\frac{G\left(  v\right)  }{1-\cos\theta
}\label{rivlinkendall}%
\end{equation}

\bigskip The case of low angles is exactly where the Kendall solution is
needed, as the Rivlin solution predicts ever increasing load and in fact
infinite loads at zero peel angles --- so Kendall is the regularized version
of the Rivlin solution, predicting instead much lower loads. Kendall made
exactly experiments at low angles, a force was applied to the film and the
crack speed was determined. The angle in the experiment was therefore adjusted
until the crack speed was $80\mu m/s$ so to correspond to an adhesive energy
of $P_{Rivlin}\left(  80\mu m/s\right)  =G\left(  80\mu m/s\right)  =5N/m$ as
determined from peeling tests at angle $\pi/2$, i.e. equation
(\ref{rivlinkendall}). Kendall verified this equation for angles as low as 10$%
{{}^\circ}%
$ where the tape was ethylene propylene rubber adhering on glass. This
verified very well Kendall's equation which therefore we can interpret already
as a "viscoelastic tape" solution as
\begin{equation}
\frac{G\left(  v\right)  }{EL}=\frac{1}{2}\left(  \frac{P\left(  v\right)
}{EL}\right)  ^{2}+\left(  \frac{P\left(  v\right)  }{EL}\right)  \left(
1-\cos\theta\right)
\end{equation}
where $G\left(  v\right)  $ is what has been measured using
(\ref{rivlinkendall}). This solution shows no critical speeds, and finds a
unique (positive) solution
\begin{equation}
\left(  \frac{P\left(  v\right)  }{EL}\right)  _{1,2}=-\left(  1-\cos
\theta\right)  +\sqrt{\left(  1-\cos\theta\right)  ^{2}+2\frac{G\left(
v\right)  }{EL}}>0
\end{equation}
so at low angles we simply have
\begin{equation}
\left(  \frac{P\left(  v\right)  }{EL}\right)  _{\theta=0}=\sqrt
{\frac{2G\left(  v\right)  }{EL}}=\sqrt{\frac{2\left[  P\left(  v\right)
\right]  _{\theta=\pi/2}}{EL}}\label{final}%
\end{equation}
where we used that using $G\left(  v\right)  $ from the peeling tests at angle
$\pi/2$, i.e. equation (\ref{rivlinkendall}) $G\left(  v\right)  =\left[
P\left(  v\right)  \right]  _{\theta=\pi/2}$ so nothing extraordinary seems to
occur at low angles with respect to toughness enhancement. On the contrary, if
toughness increases at large angles by a factor, say, of 1000, then we expect
a reduced toughness enhancement at low angles because of the square root in
(\ref{final}). This is not exactly found in Kendall's figures, since the same
range of loads is found at $\theta=\pi/2$ and $\theta=0$ in the experiments,
so this requires some further clarification. But the AC theory predicts a much
different result anyway, since the toughness enhancement at low angles is
dramatically higher than at large angles, as in results like fig.2.

\section{Conclusions}

We have discussed that the very simple "ultratough peeling" theory of
Afferrante and Carbone [1] leads to two possible equilibrium solutions for the
load at a given peeling speed. The low load solution gives in some ranges very
high toughness enhancement, possibly already much larger even than that
expected in a crack in a infinite system.\ More problematic appears the high
load solution which, even considering a limit stress carried by tape and
substrate, will not avoid some solutions to have unbounded toughness
enhancement. The viscoelastic effect seems to induce an on-off mechanism,
where for large angles of peeling it predicts solutions extremely close to the
Kendall elastic one, while only below a certain angle threshold, this
"ultratough" peeling regime appears. A qualitative discussion of the
viscoelastic experiments of Kendall has been added, and his treatment of the
viscoelastic effects (although these were relative to a viscoelastic tape, and
not substrate).

\section{Acknowledgements}

MC acknowledges support from the Italian Ministry of Education, University and
Research (MIUR) under the program "Departments of Excellence" (L.232/2016).

\section{References}

[1] Afferrante, L., Carbone, G. (2016), The ultratough peeling of elastic
tapes from viscoelastic substrates, \textit{Journal of the Mechanics and
Physics of Solids} 96, 223-234

[2] Kendall, K., (1975). Thin-film peeling -- elastic term. \textit{J. Phys.
D: Appl. Phys.} 8, 1449-- 1452.

[3] Long, R., Hui, C. Y., Gong, J. P., \& Bouchbinder, E. (2020). The fracture
of highly deformable soft materials: A tale of two length scales. arXiv
preprint arXiv:2004.03159.

[4] Creton, C., \& Ciccotti, M. (2016). Fracture and adhesion of soft
materials: a review. \textit{Reports on Progress in Physics}, 79(4), 046601.

[5] Gent, A. N. and Schultz, J., (1972), Effect of wetting liquids on the
strength of adhesion of viscoelastic material, \textit{J. Adhes.} 3(4), 281-294.

[6] Barquins, M., \& Maugis, D. (1981). Tackiness of elastomers. \textit{The
Journal of Adhesion}, 13(1), 53-65.).

\bigskip\lbrack7] Gent, A. N., \& Petrich, R. P. (1969). Adhesion of
viscoelastic materials to rigid substrates. \textit{Proceedings of the Royal
Society of London. A. Mathematical and Physical Sciences}, 310(1502), 433-448.

[8] Andrews, E. H., \& Kinloch, A. J. (1974). Mechanics of elastomeric
adhesion. In Journal of Polymer Science: Polymer Symposia (Vol. 46, No. 1, pp.
1-14). New York: Wiley Subscription Services, Inc., A Wiley Company.

[9] Barber, M., Donley, J., \& Langer, J. S. (1989). Steady-state propagation
of a crack in a viscoelastic strip. \textit{Physical Review A}, 40(1), 366.

[10] Greenwood J A and Johnson K L (1981) The mechanics of adhesion of
viscoelastic solids \textit{Phil. Mag. A} 43 697--711

[11] Williams, M. L.; Landel, R. F.; Ferry, J. D. (1955) The Temperature
Dependence of Relaxation Mechanisms in Amorphous Polymers and Other
Glass-Forming Liquids. \textit{Journal of the American Chemical Society}, 77, 3701-3707.

[12] Rice, J. R. (1978). Mechanics of quasi-static crack growth (No.
COO-3084-63; CONF-780608-3). Brown Univ., Providence, RI (USA). Div. of Engineering.

[13] Graham GAC 1969 Two extending crack problems in linear viscoelasticity
theory \textit{Q. Appl. Math.} 27 497--507

\bigskip\lbrack14] Schapery R A (1975) A theory of crack initiation and growth
in viscoelastic media \textit{Int. J. Fracture} 11 (Part I) 141--59

[15] de Gennes, P. G. (1996). Soft adhesives. \textit{Langmuir}, 12(19), 4497-4500.

\bigskip\lbrack16] Rivlin, R.S., (1944), The effective work of adhesion,
\textit{Paint Technol.} 9, 215-216.

\end{document}